\documentclass[aps,prb,twocolumn,showpacs,floatfix]{revtex4-1}
\usepackage{amssymb}
\usepackage{graphicx}

\begin{document}

\title{Oscillations of the Nernst coefficient in bismuth}

\author{Yu.~V.~Sharlai}
\affiliation{B.~Verkin Institute for Low Temperature Physics \&
Engineering, Ukrainian Academy of Sciences, Kharkov 61103,
Ukraine}

\author{G.~P.~Mikitik}
\affiliation{B.~Verkin Institute for Low Temperature Physics \&
Engineering, Ukrainian Academy of Sciences, Kharkov 61103,
Ukraine}

\begin{abstract} We calculate the magnetic-field dependence
(oscillations) of the Nernst coefficient in bismuth at low
temperatures for the case when the magnetic field is directed
along the trigonal axis of the crystal. In the calculations we
take into account the scattering of the electrons and holes in
bismuth on impurities and the dependence of this scattering on the
magnetic field. The results of these calculations are compared with
the experimental data recently published.
\end{abstract}


\maketitle

\section{Introduction}

In a recent paper,\cite{B1} oscillations of the Nernst coefficient
in bismuth were observed for magnetic fields directed along
the trigonal and bisectrix axes of the crystal. These oscillations
have the shape of rather sharp peaks which originate from
the crossing of the Landau levels of charge carriers in bismuth
with the Fermi level $E_F$ of this semimetal. Moreover, several
unusual peaks of this coefficient were discovered for very high
magnetic fields $H$ ($14\leq H\leq 33$ T) parallel to the trigonal
axis.\cite{B2} At such magnetic fields almost all the Landau
levels are empty, and one can hardly expect that the unusual peaks
result from the above-mentioned crossing. So the authors of
Ref.~\onlinecite{B2} suggested that the unusual peaks are caused
by some collective effects in the electron system of bismuth.
Interestingly, in the same interval of high magnetic fields
several jumps of magnetization were observed, which were ascribed
to field-induced instabilities of the ground state of interacting
electrons in bismuth.\cite{Ong}

Within a simple model of the electron energy spectrum of bismuth
in magnetic fields, the {\it positions} of the usual peaks in the
magnetic-field axis were explained, \cite{SM1,AB} by calculating the
crossing of the appropriate Landau levels with $E_F$. It turned
out that most of these peaks are due to the hole Fermi surface
of bismuth, while some result from its electron Fermi surface.
However, the unusual peaks observed in the Nernst signal at high
magnetic fields cannot be explained in this way. Nevertheless, we
showed \cite{SM1} that in principle, the positions of some of
these peaks can be reproduced if one assumes that a small
deviation of the magnetic-field direction from the trigonal axis
occurred in the experiments. In order to support or to disprove
this possible explanation of the unusual peaks, we also suggested
\cite{SM1} to investigate experimental investigation of the dependence of the
Nernst-coefficient peaks on the tilt angle of the magnetic field.
These experiments were described in
Ref.~\onlinecite{KBstrangePeaks}, and it was found that the
angular dependences of the usual peaks are in agreement with the
theory,\cite{SM1,AB} while the dependences of the unusual peaks
are essentially different, and hence their origin cannot be
explained by a small deviation of the magnetic-field direction
from the trigonal axis. In addition, the results of the
calculations in Ref.~\onlinecite{SM1} show that the usual electron
peaks observed in the experiment correspond to only some of the
crossings of the electron Landau levels with $E_F$, while other
crossings do not manifest themselves.

In order to continue analysis of the experimental findings,
\cite{B1,B2,KBstrangePeaks} in this paper we calculate not only
the positions of the peaks in the Nernst coefficient but also
their {\it shape}, i.e. the dependence of this coefficient on the
magnetic field, in the case when the field is along the trigonal
axis. We carry out the calculations within the same model of the
spectrum of bismuth as in Ref.~\onlinecite{SM1}. This model is
briefly described in Sec.~II. In Sec.~III we present formulas for
the Nernst coefficient, taking into account the effect of
impurities on the density of states of charge carriers in bismuth,
and in Sec.~IV we compare the results of our calculations with the
experimental data.

\section{Energy spectrum of bismuth}

The Fermi surface of bismuth consists of one hole ellipsoid
located at the point T of its Brillouin zone and three electron
ellipsoids centered at the points L and elongated along the
bisectrix directions. \cite{Ed} As in Ref.~\onlinecite{SM1}, we
shall describe the electron {\it spectrum of bismuth in the
magnetic field} ${\bf H}$ by the model of Smith, Baraff and
Rowell. \cite{Sm} According to this model, the $n$th Landau level
$\varepsilon_n$ for an electron with the quasimomentum $p_H$ along
${\bf H}$ can be found from the equation:
\begin{equation}\label{1}
E\left (1+{E\over E_g}\right ) = (n+{1\over 2})\hbar \omega_c
+{p_H^2\over 2m_H}\pm {1\over 2}g\beta_0H,
\end{equation}
where the signs $\pm$ correspond to the electron spins that are
antiparallel and parallel to ${\bf H}$, respectively; the energy
$E$ is measured from the edge of the conduction band; $\omega_c$
is the cyclotron frequency
\[
\omega_c={|e|H\over m_c c},
\]
$E_g$ is the gap between the conduction and valence bands at the
point L; $g$ is the effective electron g factor at this point;
$\beta_0$ is the Bohr magneton; $c$ is the speed of light; $e$ is
the electron charge, and the longitudinal and the cyclotron
masses, $m_H$ and $m_c$, are given by
\begin{eqnarray}\label{2}
 m_H&=& {\bf h}\cdot {\bf m}^e\cdot {\bf h}, \\
 m_c&=&[\det{\bf m}^e/m_H]^{1/2}. \label{3}
\end{eqnarray}
Here ${\bf h}$ is the unit vector in the direction of the magnetic
field ${\bf H}$. The effective mass tensor ${\bf m}^e$ has the
form:
\begin{equation}\label{4}
{\bf m}^e= \left (
    \begin{array}{ccc}
    m_{11} & 0 & 0 \\
    0 & m_{22} & m_{23} \\
    0 & m_{23} & m_{33}
    \end{array}
\right ),
\end{equation}
where the axes 1 and 3 coincide with the binary and the trigonal
axes, respectively, while the axis 2 is along the bisectrix
direction. The effective $g$ factor,
\begin{equation}\label{5}
g^2= 4m^2{{\bf h}\cdot {\bf m}^e_s\cdot {\bf h}\over \det{\bf
m}^e_s },
\end{equation}
is defined in terms of a spin-mass tensor ${\bf m}^e_s$ that has
the form similar to Eq.~(\ref{4}). Within the model of Smith,
Baraff, and Rowell \cite{Sm} the elements of ${\bf m}^e_s$ differ
from the elements of ${\bf m}^e$.

\begin{table}
\caption{\label{table1} Parameters of the Smith-Baraff-Rowell
spectrum\cite{Sm}}
\begin{ruledtabular}
\begin{tabular}{lcccc}
Electrons&$m_{11}$&$m_{22}$&$m_{33}$&$m_{23}$\\ \hline Orbital
mass & 0.00113 & 0.26 & 0.00443 & -0.0195\\ Spin mass & 0.00101 &
2.12 & 0.0109 & -0.13\\ \hline Holes&&$M_1=M_2$&&$M_3$\\ \hline
Orbital mass&&0.07&&0.69\\ Spin mass&&0.033&&200\\ \hline
$E_g=15.3\,{\rm meV}$ &&& $E_0=38.5\,{\rm meV}$
\end{tabular}
\end{ruledtabular}
\end{table}

Since the spectrum of the holes at the point T is parabolic, their
Landau levels can be easily found \cite{Ed}
\begin{equation}\label{6}
E_0- E = (n+{1\over 2})\hbar \omega_c +{p_H^2\over 2m_H}\pm
{1\over 2}g\beta_0H,
\end{equation}
where $E_0$ is the edge of the hole band at this point of the
Brilloun zone. The cyclotron frequency $\omega_c$, the masses
$m_c$ and $m_H$, and the $g$ factor are defined by the same
formulas (\ref{2}), (\ref{3}), (\ref{5}) as for the electrons, but
now the tensor of the effective masses for the holes ${\bf m}^h$
has the form:
\begin{equation}\label{7}
{\bf m}^h= \left (
    \begin{array}{ccc}
    M_1 & 0 & 0 \\
    0 & M_2 & 0 \\
    0 & 0 & M_3
    \end{array}
\right ),
\end{equation}
and a similar expression is valid for the spin-mass tensor, ${\bf
m}_s^h$. All the parameters in Eqs.~(\ref{1}) - (\ref{7}) are
known for bismuth; \cite{Sm,ex1} see Table \ref{table1}.

The above formulas lead to the following expression for the
contribution of the i-th electron ellipsoid to the density of
electrons in bismuth, \cite{Sm}
\begin{equation}\label{Ne}
N_i^{e}(E_F,H)=\frac{(2m_H)^{1/2}eH}{2\pi^2\hbar^2c}
\sum_{n,\pm}\sqrt{E_F^{*}-E_{n,\pm}},
\end{equation}
where
\[
E_F^{*}=E_F(1+E_F/E_g),
\]
$E_F$ is the Fermi energy, and we have used the notation
\[
E_{n,\pm}=(n+1/2)\hbar\omega_c \pm (1/2)\beta_0 g H .
\]
The density of the holes is given by \cite{Sm}
\begin{equation}\label{Nh}
N^h(E_F,H)=\frac{(2m_H)^{1/2} eH}{2\pi^2\hbar^2c}\sum_{n,\pm}
 \sqrt{E_0-E_F-E_{n,\pm}}.
\end{equation}
The position of the Fermi energy $E_F(H)$ is determined by the
condition of charge neutrality:
\begin{equation}\label{Z=0}
Z\equiv \sum_{i=1}^{3}N_i^{e}(E_F,H)-N^h(E_F,H)=0.
\end{equation}
Below we shall apply the formulas of this section to the case when
the magnetic field is directed along the trigonal axis of the
crystal, i.e., when ${\bf h}=(0,0,1)$.

\section{Nernst coefficient}

Let the $z$ axis of the coordinate system be directed along the
trigonal axis of the crystal, while the $x$ and $y$ axes lie in
the basal plane of bismuth. In the presence of a weak electric
field ${\bf E}$ and of a small temperature gradient ${\bf
\nabla}T$, the charge-current density ${\bf j}$ is determined by
\begin{equation}\label{jET}
j_i=\sigma_{ik}E_{k}^{*}-\alpha_{ik} \nabla_{k}T ,
\end{equation}
where the subscripts $i$, $k$ stand for the spatial coordinates
$x$,$y$,$z$; ${\bf E}^{*}={\bf E}+\frac{1}{e}\nabla\mu$; $\mu$ is
the local chemical potential in the crystal; $\sigma_{ik}$ and
$\alpha_{ik}$ are the conductivity and Peltier (thermoelectric)
tensors, respectively. In the absence of an external magnetic field
the tensors $\sigma_{ik}$ and $\alpha_{ik}$ are diagonal, and one
has
\begin{equation}\label{xx}
\sigma_{xx}=\sigma_{yy}, \ \ \ \alpha_{xx}=\alpha_{yy}.
\end{equation}
If the external magnetic field ${\bf H}$ is applied along the $z$
axis, the off-diagonal terms $\sigma_{xy}$, $\sigma_{yx}$ and
$\alpha_{xy}$, $\alpha_{yx}$ become different from zero, and they
satisfy the relationships:
\begin{equation}\label{xy}
\sigma_{xy}=-\sigma_{yx}, \ \ \  \alpha_{xy}=-\alpha_{yx} .
\end{equation}
In this situation the temperature gradient $\nabla_x T$ will
generate a transverse electric field $E_y$. The Nernst
coefficient $\nu$ is defined by the relation
\begin{equation}\label{nu}
\nu H =|\frac{E_{y}}{\nabla_{x}T}|_{{\bf j}=0} ,
\end{equation}
and it can be found from the formula
\begin{equation}\label{N}
S_{xy}\equiv \nu H=|\frac{\sigma_{xx}\alpha_{xy}-
\sigma_{xy}\alpha_{xx}}{\sigma_{xx}^{2}+\sigma_{xy}^{2}}|,
\end{equation}
which immediately follows from Eqs.~(\ref{jET})--(\ref{nu}).

Since the oscillations of the Nernst coefficient can be observed
only at sufficiently high magnetic fields, we assume below that
the condition
\begin{eqnarray}\label{conditionOmegaTau}
 \omega_c \tau \gg 1,
\end{eqnarray}
is fulfilled for both the electrons and the holes in bismuth where
$\tau$ is the scattering time of an electron or a hole on
impurities. In clean crystals of bismuth condition
(\ref{conditionOmegaTau}) can be satisfied even for moderate
magnetic fields. For example, using the parameters presented in
Table I and the experimental data of Ref.~\onlinecite{KBlowField},
we estimate that at $H=1$ kOe directed along the trigonal axis the
values of $\omega_c \tau$ for the electrons and holes are of
the order of $10$ for the crystal used by K. Behnia {\it et
al}.\cite{KBlowField} (here we have assumed that the scattering
lengths for the holes and the electrons are approximately the
same).

Under condition (\ref{conditionOmegaTau}), the diagonal component
$\alpha_{xx}$ of the Peltier tensor is small as compared with
$\alpha_{xy}$:\cite{kappaXX}
\begin{equation}
\frac{\alpha_{xx}}{\alpha_{xy}} \sim \frac 1{\omega_c \tau}\ll 1 .
\end{equation}
Since bismuth has equal numbers of electrons and holes, the
ratio $\sigma_{xy}/\sigma_{xx}$ in this semimetal is also
suppressed as compared to the case of a metal with a single type
of charge carrier \cite{Abr} (e.g., in
Ref.~\onlinecite{KBHallEffect} $\sigma_{xx}$ was approximately ten
times larger than $\sigma_{xy}$). Then, we obtain from
Eq.~(\ref{N}) that
\begin{equation} \label{Ner}
\nu H\approx \frac{\sigma_{xx}\alpha_{xy}}{\sigma_{xx}^{2}+
\sigma_{xy}^{2}}=\rho_{xx}\alpha_{xy},
\end{equation}
where $\rho_{ik}=\sigma_{ik}^{-1}$ is the resistivity tensor. As
the experimental data \cite{KBrhoXX} show, $\rho_{xx}$ is a smooth
and almost monotonic function of the magnetic field in that region
of the $T-H$ plane in which the oscillations of the Nernst coefficient
were observed. \cite{B1,B2} Hence, the peaks in the Nernst
coefficient mainly result from the peaks in $\alpha_{xy}$, and
below we shall analyze only this $\alpha_{xy}$, taking
$\rho_{xx}(H)$ from the experiment. \cite{KBrhoXX}

The off-diagonal term of the Peltier tensor, $\alpha_{xy}$,
consists of the purely electronic and the phonon-drag parts,
\begin{equation}\label{EqKappaSum}
\alpha_{xy}=\alpha_{xy}^{(e)}+ \alpha_{xy}^{(ph)}.
\end{equation}
The purely electronic part $\alpha_{xy}^{(e)}$ is caused by a
nonequilibrium distribution of the electrons and holes over the
crystal when $\nabla_x T \neq 0$. In strong magnetic fields,
$\omega_c\tau \gg 1$, $\alpha_{xy}^{(e)}$ is dissipationless, i.e.
it does not depend on the relaxation mechanisms of this distribution:
\cite{obr,BergmanOganesyan}
\begin{equation}\label{kappaAprox2}
\alpha_{xy}^{(e)}\!=\!\frac{cS}H\!=\!
\frac{c}{H}\!\int_{-\infty}^{+\infty}\!\!\!\!\!\!\!\!dE
\frac{\partial Z}{\partial E}\{f_F\ln f_F+(1-f_F)\ln (1-f_F)\},
\end{equation}
where $S$ is the entropy of the charge carriers per a unit volume;
$f_F$ is the Fermi function with the chemical potential $\mu$, and
$\partial Z/\partial E$ is the density of states of the charge
carriers in bismuth,
\begin{equation}\label{Z}
\frac{\partial Z(E,H)}{\partial E} = \sum_{i=1}^3 \frac{\partial
N_i^{e}(E,H)}{\partial E} \\ +\left | \frac{\partial
N^h(E,H)}{\partial E}\right |.
\end{equation}
According to formulas (\ref{1}), (\ref{6}), (\ref{Ne}), and
(\ref{Nh}), the derivative $\partial Z/\partial E$ diverges when
the energy $E$ crosses the Landau levels. The integral over $E$
with the hat-like function $\{f_F\ln f_F+(1-f_F)\ln (1-f_F)\}$ of
characteristic width $T$ smooths the singularities of
$\partial Z/\partial E$, but at low temperatures, $\hbar\omega_c
\gg T$, the quantity $\alpha_{xy}^{(e)}$ shows sharp peaks
when the Fermi energy approaches a Landau level. These simple
considerations qualitatively explain the origin of the peaks in
the Nernst coefficient of bismuth.

The phonon-drag part $\alpha_{xy}^{(ph)}$ originates from the
charge carriers that are dragged by a phonon flow in the crystal
due to the electron-phonon interaction. In the strong magnetic
fields this part is described by a sufficiently complicated
formula which schematically has the form:
\cite{phononDrag,JayGerin}
 \begin{equation}\label{phd}
\alpha_{xy}^{(ph)}\propto \frac{c}{H}\left\langle\frac{\nu_{pe}}
{\nu_{pe}+\nu_0}\right\rangle,
 \end{equation}
where $\nu_{pe}$ is the probability of phonon scattering by the
charge carriers; $\nu_0$ is the probability of phonon scattering
by other phonons and by the boundaries of the sample; $\langle
\dots \rangle$ means some averaging over the wave vectors of the
phonons existing at the temperature $T$. It is essential here that
the probability $\nu_{pe}$ is proportional to the density of
states $\partial Z/\partial E$ and oscillates with changing $H$.
At very low temperatures the phonon scattering by the boundaries
of the sample prevails over other phonon scattering processes,
\cite{KBlowField} i.e., one has $\nu_0 \gg \nu_{pe}$. In this case
the expression for $\alpha_{xy}^{(ph)}$ is proportional to
$\partial Z/\partial E$ and is similar to Eq.~(\ref{kappaAprox2}).
In order to estimate the magnetic-field dependence of the phonon-drag
part in this situation, we shall use the simplest approximation
 \begin{equation}\label{ph}
\alpha_{xy}^{(ph)}(H,T)\approx r(T)\alpha_{xy}^{(e)}(H,T),
 \end{equation}
where the factor $r(T)$ can be found as the ratio of
$\alpha_{xy}^{(ph)}$ and $\alpha_{xy}^{(e)}$  at sufficiently low
magnetic fields when the oscillations of $\alpha_{xy}$ become
small. With increasing $T$ the probability $\nu_{pe}$ may become
of the order of $\nu_0$ or even exceed it. According to
Ref.~\onlinecite{KBlowField}, this seems to occur in bismuth at
$T\sim 3-4$ K. Then, $\nu_{pe}/(\nu_{pe}+\nu_0)\sim 1$, and the
oscillations of $\partial Z/\partial E$ with changing $H$
scarcely reveal themselves in $\alpha_{xy}^{(ph)}$.

\subsection{Effect of impurities on the electron and hole densities
of states in bismuth}

Although in the limit of strong magnetic fields, $\omega_c
\tau \gg 1$, formula (\ref{kappaAprox2}) does not depend on $\tau$
explicitly, the scattering of the charge carriers on
impurities renormalizes the density of states of the electrons and
holes in bismuth, causing a smearing of the peaks in
$\alpha_{xy}$. We calculate the renormalized $\partial Z/\partial
E$ for bismuth using the approach of Ref.~\onlinecite{Bychkov}.

Within  this approach  Green functions $G_l$ of charge carriers
in the magnetic field $H$ are considered that are averaged over
the positions of impurities in the crystal. The index $l$ means the
set of quantum numbers $n$, $p_z$, $p_y$, $s_z=\pm1$ that
characterize eigenstates of a charged particle in a magnetic
field in the absence of impurities. The potential of an impurity,
$U({\bf r})$, is assumed to be of the form $U({\bf r})=f
\delta({\bf r})$ where $\delta({\bf r})$ is the delta function,
and the constant factor $f$ determines the scattering amplitude of
a charged particle on an impurity at $\varepsilon=0$ and $H=0$. The
density of states of the charge carriers per unit interval of
their energy $\varepsilon$, $\partial N(\varepsilon)
/\partial\varepsilon$, is expressed in terms of the Green
functions as
\begin{equation}\label{G}
\frac{\partial N}{\partial \varepsilon}=\frac{1}{\pi}{\rm Im}
G(\varepsilon),
\end{equation}
where $G(\varepsilon)=\sum_lG_l(\varepsilon)$. The function
$G(\varepsilon)$ can be found from the self-consistent
equation\cite{Bychkov}
\begin{equation}\label{eqOnG}
G=\sum_{l}(\varepsilon_l-\varepsilon-c_if^{2}G)^{-1},
\end{equation}
where $\varepsilon_l =\varepsilon(n,p_z,s_z)$ is the spectrum of a
charge carrier in the magnetic field in the absence of impurities, and
$c_i$ is the impurity concentration. The quantity $f$ can be
expressed in terms of the scattering time $\tau$ of the electrons
(the holes) at $H=0$ when the energy $\varepsilon$ of these
particles is equal to the chemical potential $\mu_0$ of bismuth in
zero magnetic field:
\begin{equation}
c_if^{2}G|_{H\rightarrow0,\varepsilon\rightarrow\mu_{0}}=
i\frac{\pi\hbar}{2\tau}.
\end{equation}
Below we calculate $G$ under the simplification that the
scattering of the holes and of the electrons of the three
different ellipsoids of bismuth does not mix these particles, and
so one has electron $\tau_e$ and hole $\tau_h$ scattering
times.

The integration over $p_z$ and $p_y$ in Eq.~(\ref{eqOnG}) can be
done explicitly for the electrons and holes that are described by
the spectra (\ref{1}) and (\ref{6}), respectively. Then, in the
case of the holes we obtain the following equation in $G$:
\begin{equation}\label{eqOnG2}
G-G_0=i\frac{m_z^{1/2}eH}{2^{5/2}\pi\hbar^2c}\sum_{n,\pm}\left \{
\\ g_{n,\pm}(c_if^2G)-g_{n,\pm}(0) \right \},
\end{equation}
where
\begin{equation}
g_{n,\pm}(x)=\frac1{\sqrt{E_0-\varepsilon+x-E_{n,\pm}}}\,,
\end{equation}
and  $G_0(\varepsilon)$ is the function $G(\varepsilon)$ in
the absence of impurities. The function $G_0$ is imaginary and is
given by
\begin{equation}
{\rm Im}G_0=\pi \frac{\partial N_0^h(\varepsilon,H)}{\partial
\varepsilon},
\end{equation}
where $\partial N_0^h/\partial \varepsilon$ is the density of
states of the holes in the absence of impurities, i.e, $\partial
N_0^h/\partial \varepsilon$ directly follows from Eq.~(\ref{Nh}).
The functions $G$ for all the electron groups are identical, and
the equation in $G$ has the same form as Eq.~(\ref{eqOnG2}) but
with
\begin{equation}
g_{n,\pm}(x)=\frac{E_g+2(\varepsilon+x)}{E_g^{1/2}\left[
(\varepsilon+x)^2+E_g(\varepsilon+x-E_{n,\pm})\right]^{1/2}},
\end{equation}
and
\begin{equation}\label{G0}
{\rm Im}G_0=\pi \frac{\partial N_0^e(\varepsilon,H)}{\partial
\varepsilon},
\end{equation}
where $\partial N_0^e/\partial \varepsilon$ is the density of
states of one of the electron ellipsoids in absence of the
impurities. The renormalized densities of states of the electrons
and of the holes in bismuth calculated with equations (\ref{G})--
(\ref{G0}) are shown in Figs.~1 and 2.

\begin{figure}
\includegraphics[scale=1]{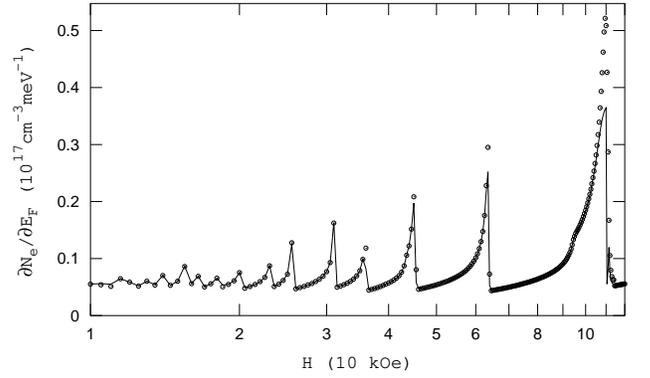}
\caption{\label{Graph6} The density of states for the electrons in
bismuth. The solid line has been calculated with Eqs.~(\ref{G})--
(\ref{G0}) and $\hbar/\tau_e=0.001$ meV. The points show the
analytical result obtained with the substitution (\ref{sub}) at
$\hbar/\tau_e'=0.1$ meV.
 } \end{figure}

 \begin{figure}
\includegraphics[scale=1]{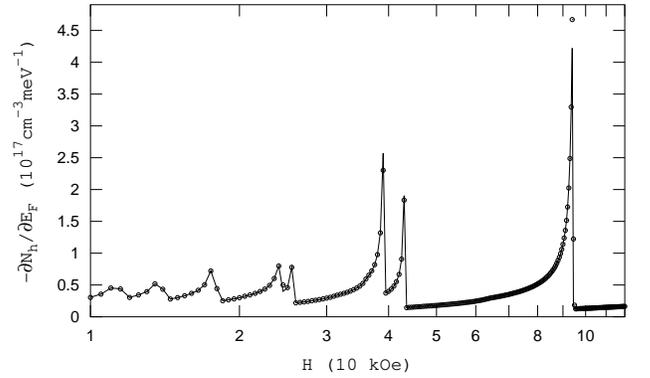}
\caption{\label{Graph7}The density of states for the holes in
bismuth. The solid line has been calculated with Eqs.~(\ref{G})--
(\ref{G0}) and $\hbar/\tau_h=0.5\cdot 10^{-3}$ meV. The points
show the analytical result obtained with the substitution
(\ref{sub}) at $\hbar/\tau_h'=0.007$ meV.
 } \end{figure}

Interestingly, the renormalized densities of states can be
approximately obtained by the following simple substitution in
formulas (\ref{Ne}), (\ref{Nh}):
\begin{equation}\label{sub}
u \rightarrow  {\sqrt{u^2+\Gamma^2}+u \over 2},
\end{equation}
and by the subsequent differentiation of these formulas over
$E_F$. Here $\Gamma = \hbar/\tau'$, $u=E_0-E_F-E_{n,\pm}$ for the
holes and $u=E_F^{*}-E_{n,\pm}$ for the electrons. It turns out
that at an appropriate choice of the effective scattering times
$\tau'$, the densities of state thus obtained very well describe
the densities of states calculated with Eqs.~(\ref{G})--
(\ref{G0}), see Figs.~1 and 2. The effective scattering time
$\tau'$ weakly depends on the number $n$ of the Landau level
nearest to $E_F$, and so one and the same value of $\tau'$ can
describe the oscillations of the density of states in a rather
wide interval of the magnetic fields, Figs.~1 and 2.
Note that $\tau'$ found for strong magnetic fields is essentially
larger than $\tau$ defined in the limit $H\to 0$.

The substitution (\ref{sub}) can be understood from the following
considerations: Let us average the density of states of charge
carriers in the absence of impurities, $\partial N_0(E)/\partial E$,
with a hat-like function characterized by some half-width
$\Gamma=\hbar/\tau'$:
\begin{equation}
{\partial N(\varepsilon)\over \partial \varepsilon}=\frac 1\pi
\int_{-\infty}^\infty dE
\frac{(\hbar/\tau')}{(E-\varepsilon)^2+(\hbar/\tau')^2}{\partial
 N_0(E)\over \partial E}.
\end{equation}
In the case of the holes in bismuth (and for any
quadratic spectrum) this integral is calculated analytically, and
the density of states obtained coincides with $\partial
N(E_F)/\partial E_F$ calculated with the substitution (\ref{sub}).

Using the substitution, one can easily understand the result that
follows from the calculations with Eqs.~(\ref{G})-- (\ref{G0}): In
the presence of impurities, the positions of the peaks in $\partial
N(E_F)/\partial E_F$ generally do not coincide with the magnetic
fields at which the Landau levels calculated at zero impurity
concentration, $c_i=0$, cross the Fermi energy. Indeed, for the
holes in bismuth, we find from formula (\ref{Nh}) with
substitution (\ref{sub}) that
 \begin{equation}\label{eqDNhDE2}
\left | \frac{\partial N^h}{\partial E_F}\right |\approx
\frac{(2m_z)^{1/2}eH}{4\sqrt{2}\pi^2\hbar^2c}\sum_{n,\pm} \\ \frac
{\sqrt{\sqrt{u_{n,\pm}^2+\Gamma^2}+
u_{n,\pm}}}{\sqrt{u_{n,\pm}^2+\Gamma^2}},
\end{equation}
where $u_{n,\pm}=E_0-E_F-E_{n,\pm}$. Now each term of this sum
reaches its maximum at $u_{n,\pm}=\Gamma/\sqrt 3$. On the other
hand, in the absence of impurities the singularities in $\partial
N_0(E_F)/\partial E_F$ occur at $u_{n,\pm}=0$. In other words, if
one takes into account the impurities, the positions of the peaks
in the density of states shift.

\section{Results}

\begin{figure}
\includegraphics[scale=1]{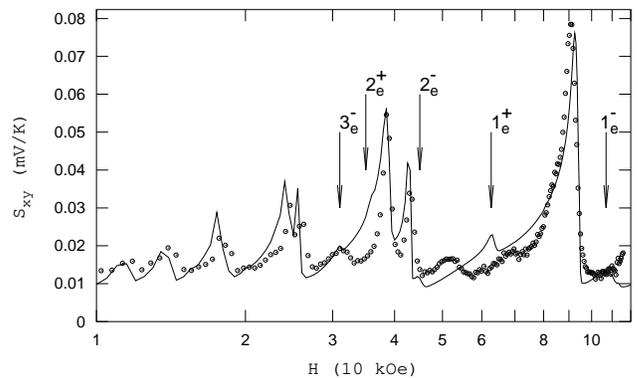}
\caption{\label{Sxy018} The Nernst signal $S_{xy}=\nu H$ as a
function of the magnetic field, $H$, at $T=0.18$ K. The points
show the experimental data, \cite{KBstrangePeaks} and the solid
line is the result of the calculation with Eqs.~(\ref{Ner}) --
(\ref{G0}). Here $r(0.18)=0.85$, $\hbar/\tau_h = 0.005 {\rm meV}$,
and $\hbar/\tau_e = 0.005 {\rm meV}$. The arrows indicate
positions of the electron peaks visible in the solid line.
 } \end{figure}

\begin{figure}
\includegraphics[scale=1]{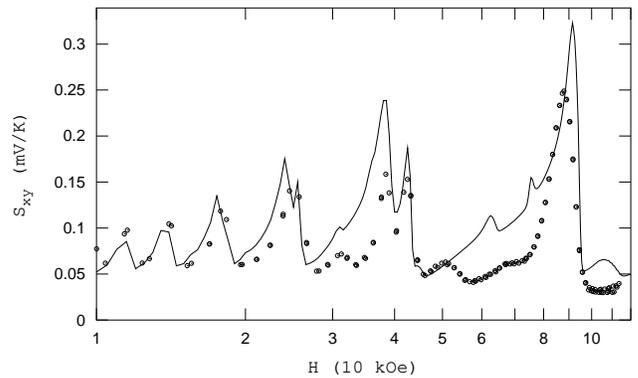}
\caption{\label{Sxy049} As Fig.~\ref{Sxy018} but here $T=0.49$ K
and $r(0.49)=2.5$.
 } \end{figure}

\begin{figure}
\includegraphics[scale=1]{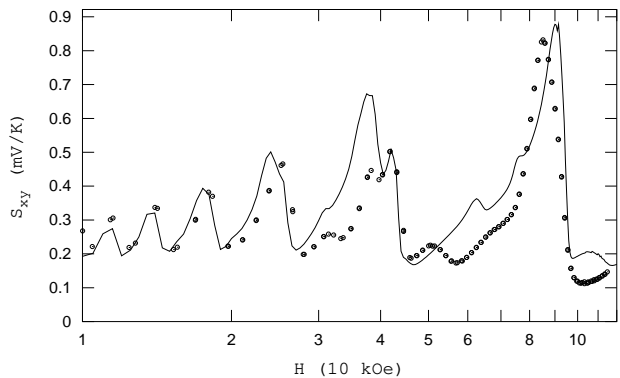}
\caption{\label{Sxy104} As Fig.~\ref{Sxy018} but here $T=1.04$ K
and $r(1.04)=4.5$.
 } \end{figure}

\begin{figure}
\includegraphics[scale=1]{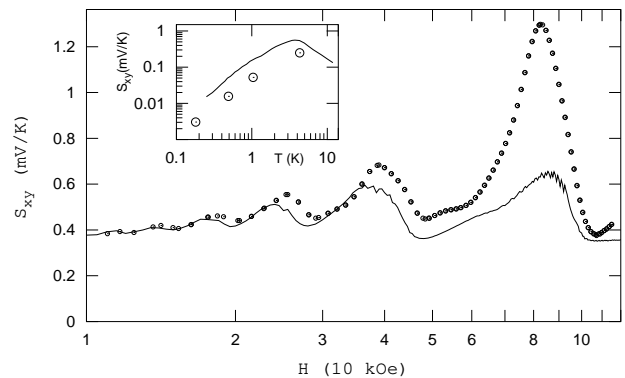}
\caption{\label{Sxy43}As Fig.~\ref{Sxy018} but here $T=4.3$ K and
instead of formula (\ref{ph}) we use $S_{xy}^{(ph)}(H)=0.21$ mV/K.
The inset shows the function $S_{xy}(T)$ at $H=1$ kOe (the solid
line represents the experimental data, \cite{KBlowField} while the
points are the values calculated here).
 } \end{figure}

In Figs.~\ref{Sxy018}--\ref{Sxy43} we show the magnetic-field
dependences of the quantity $S_{xy}=\nu H$ calculated with
Eqs.~(\ref{Ner})-- (\ref{G0}) at different temperatures $T$, using
the data on the resistivity $\rho_{xx}(H)$ measured in Ref.~
\onlinecite{KBrhoXX}. For comparison, in these figures we also
present the appropriate experimental data \cite{KBstrangePeaks}
for $S_{xy}(H)$. In the calculation of the curves $S_{xy}(H)$ we
take into account the $H$-dependence of the chemical potential $\mu$
which is determined by Eq.~(\ref{Z=0}). Since the renormalization
of the electron and hole densities of states by impurities
influences this dependence, we have calculated $N^h$ and $N^e_i$
in equation (\ref{Z=0}) using formulas (\ref{Ne}), (\ref{Nh}) and
substitution (\ref{sub}). At temperatures $T=0.18$ K, $0.49$ K and
$1.04$ K the factor $r$ in Eq.~(\ref{ph}) has been chosen in such
a way that the calculated and the experimental curves coincide at
a point near the lowest magnetic field $H\approx 10$ kOe, i.e. at
the point where the oscillations of $S_{xy}(H)$ are small. At
$T=4.3$ K (Fig.~\ref{Sxy43}), in accordance with the
considerations of Sec.~III, we take $S_{xy}^{(ph)}(H)$ as a
constant. The value of this constant, $S_{xy}^{(ph)}=0.21$ mV/K,
has been found from the condition that the calculated $S_{xy}$ at
$H=10$ kOe  fits the appropriate experimental value.

An inspection of Figs.~\ref{Sxy018}--\ref{Sxy43} shows that the
theoretical curves sufficiently well describe the experimental
data especially in the low-field part of the figures where the
electron peaks essentially do not manifest themselves. Note also
that the values $\tau_h=\tau_e\approx 1.2\cdot 10^{-10}$ sec used
in the construction of these figures are of the same order of
magnitude as those estimated in Ref.~\onlinecite{Hartman} from the
data of transport measurements ($\tau_h\approx 8.5\cdot 10^{-10}$
sec and $\tau_e \approx 4.4\cdot 10^{-10}$ sec). Interestingly,
although $\tau_e$ is equal to $\tau_h$ here, we find a relatively
small amplitude for the electron peaks since the hole density of
states is essentially larger than the electron contribution to
$\partial Z(E_F)/\partial E_F$. That is why only a few electron
peaks are seen in the experiment, \cite{KBstrangePeaks} and the
value of $\tau_e$ has a weak effect on the calculated $S_{xy}$.
The visible difference in the positions of the electron peaks in
the theoretical and experimental curves seems to be caused by an
insufficient accuracy of the spectrum of Smith {\it et
al}.\cite{Sm} used here. As to the discrepancy between the
hole peaks in the theoretical and experimental curves in the
high-field part of the figures, it could be due to the two reasons.
First, our simplest approximation for the phonon-drag part
$\alpha_{xy}^{(ph)}$, Eq.~(\ref{ph}), and for $S_{xy}^{(ph)}(T=4.3
{\rm K})$ [i.e., $S_{xy}^{(ph)}(T=4.3 {\rm K})=0.21$ mV/K] is
probably impaired as magnetic field $H$ deviates farther and
farther away from the point $10$ kOe at which the factors $r$ and
the value of $S_{xy}^{(ph)}(T=4.3 {\rm K})$ were found. Second,
equation (\ref{eqOnG}) obtained in Ref.~\onlinecite{Bychkov},
strictly speaking, is valid in the semiclassical limit (when the
Landau-level number $n\gg 1$), while we use this equation in the
whole interval of the magnetic fields. In other words, in the
ultraquantum limit ($n\sim 1$) a special analysis of $S_{xy}$ is
required.

In the inset of Fig.~\ref{Sxy43} we compare the values of $S_{xy}$
calculated at $H=1$ kOe for $T=0.18$ K, $0.49$ K, $1.04$ K, and
$4.3$ K with the experimental temperature dependence of $S_{xy}$
measured in Ref.~\onlinecite{KBlowField} at the same magnetic
field. These two functions $S_{xy}(T)$ are seen to agree
reasonably well if it is remembered that they correspond to
different crystals of bismuth [the phonon-drag part
$S_{xy}^{(ph)}$ depends on the dimensions of the sample at low
temperatures when the phonon mean-free path $l\propto (\nu_{pe}+
\nu_0)^{-1}$ is set by the sample size].

The  values of the factor $r$ obtained (see the captions to
Figs.~~\ref{Sxy018}--\ref{Sxy104}) demonstrate that the relative
contribution of the phonon-drag part $S_{xy}^{(ph)}$ to $S_{xy}$
decreases with decreasing $T$. Our estimates show that at $T\le
0.05$ K the phonon-drag part will practically die out, and one
will be able to neglect $S_{xy}^{(ph)}$ completely in the
calculations and hence to compare directly the electronic part
$S_{xy}^{(e)}(H)$ with the experimental data. Therefore, the
measurements of the Nernst signal at such low temperatures could
provide the possibility to compare the theory and the experiment more
accurately.

\section{Conclusions}

We calculate the purely electronic part $S_{xy}^{(e)}(H)$ of the
Nernst signal in bismuth at low temperatures for the case when the
magnetic field is directed along the trigonal axis of the crystal.
In this calculation we take into account the renormalization of
the densities of states of the electrons and holes in bismuth due
to scattering of these charge carriers on impurities. Without this
renormalization, the amplitude of the oscillations of
$S_{xy}^{(e)}(H)$ at low temperatures would be too large as
compared with the amplitude observed in the experiment.
\cite{KBstrangePeaks} As to the phonon-drag part
$S_{xy}^{(ph)}(H)$ of the Nernst signal, this part has been taken
into account within the simplest approximation. The results of the
calculations of $S_{xy}=S_{xy}^{(e)}+S_{xy}^{(ph)}$ are compared
with the experimental data, \cite{KBstrangePeaks,KBlowField} and
we find a reasonable agreement between the theoretical and
experimental curves in the region of the usual peaks of the Nernst
coefficient ($H\le 120$ kOe). We also explain the lesser importance
of the electrons than the holes in the production  of the peaks and
point out that measurements of the Nernst signal at lower
temperatures would be useful for a more accurate comparison of the
theory and the experiment. The approach of this paper, in
principle, can  also be used in analyzing the Nernst-coefficient
oscillations recently observed in graphite. \cite{graphite}

\begin{acknowledgments}

We thank K.~Behnia for useful discussions and for providing us
with the experimental data. This work was supported by the
France-Ukraine program of scientific collaboration ``DNIPRO''.
\end{acknowledgments}


\begin{thebibliography}{}

\bibitem{B1} K. Behnia, M.-A. M\'easson, and Y. Kopelevich, \prl{\bf 98}, 166602 (2007).

\bibitem{B2} K. Behnia, L. Balicas, and Y. Kopelevich,  Science
{\bf 317}, 1729 (2007).

\bibitem{Ong} Lu Li, J.~G.~Checkelsky, Y.~S.~Hor, C.~Uher,
A.~F.~Hebard, R.~J.~Cava, N.~P.~Ong, Science {\bf 321}, 547
(2008).

\bibitem{SM1} Yu. V. Sharlai and G. P. Mikitik, \prb{\bf 79}, 081102(R)
(2009).

\bibitem{AB} J. Alicea and L. Balents, \prb{\bf79}, 241101 (2009)

\bibitem{KBstrangePeaks} H. Yang, B. Fauqu\'e, L. Malone, A.B. Antunes, Z. Zhu, C. Uher, and K. Behnia, Nature Commun.  {\bf 1}, 47 (2010).

\bibitem{Ed} V.S. Edel'man, Usp. Fiz. Nauk {\bf 123}, 257 (1977) [Sov.
Phys. Usp. {\bf 20}, 819 (1977)].

\bibitem{Sm} G.E. Smith, G.A. Baraff, and J.W. Rowell, Phys. Rev. {\bf
135}, A1118 (1964).

\bibitem{ex1} Here we use the value $0.07$ for $M_1=M_2$ instead of
$0.064$ presented initially in Ref.~\onlinecite{Sm}. This value
better describes the oscillations of resistivity $\rho_{xx}(H)$
measured by S.~G.~Bompadre, C.~Biagini, D.~Maslov, and
A.~F.~Hebard, \prb{\bf 64}, 073103 (2001).

\bibitem{KBlowField} K. Behnia, M.-A. Measson, and Y. Kopelevich,
\prl{\bf 98}, 076603 (2007).

\bibitem{kappaXX} B.M. Askerov, {\it Electron transport phenomena
in Semiconductors} (World Scientific, Singapore, 1994).

\bibitem{Abr} A.A. Abrikosov, {\it Fundamentals of the Theory of
Metals} (North Holland, Amsterdam, 1988).

\bibitem{KBHallEffect} B. Fauque, H. Yang, I. Sheikin, L. Balicas,
J.-P. Issi, and K. Behnia \prb{\bf 79}, 245124 (2009).

\bibitem{KBrhoXX} B. Fauque, B. Vignolle, C. Proust, J.-P. Issi,
and K. Behnia, New J. Phys. {\bf 11} 113012 (2009).

\bibitem{obr} Yu.N. Obraztsov, Fiz. Tverd. Tela (Leningrad) {\bf 7}, 573 (1965) [Sov. Phys. Solid State {\bf 7}, 455 (1965)].

\bibitem{BergmanOganesyan}D. L. Bergman and V. Oganesyan,
Phys. Rev. Lett. {\bf 104}, 066601 (2010).

\bibitem{phononDrag} P. S. Zyryanov and G. I. Guseva, Usp. Fiz.
Nauk {\bf 95}, 565 (1968) [Sov. Phys.-Uspekhi {\bf 11}, 538
(1969)].

\bibitem{JayGerin} J.P.~Jay-Gerin, \prb{\bf 12}, 1418 (1975).

\bibitem{Bychkov} Yu. A. Bychkov, Zh. Eksp. Teor. Fiz. {\bf 39},
1401 (1960){[}Sov. Phys. JETP {\bf 12}, 977 (1961){]}.

\bibitem{Hartman} R. Hartman, Phys.~Rev. {\bf 181}, 1070 (1969).

\bibitem{graphite}Z. Zhu, H. Yang, B. Fauque, Y. Kopelevich, and
K. Behnia, Nature Physics {\bf 6}, 26 (2010).

\end{thebibliography}
\end{document}